\documentclass[sigconf, authorversion]{acmart}
\renewcommand\footnotetextcopyrightpermission[1]{}
\begin{document}

\title{HDR Lighting Dilation for Dynamic Range Reduction on \\ Virtual Production Stages}


\author{Paul Debevec}
\affiliation{%
  \institution{Netflix}
  \country{USA}
}
\email{debevec@netflix.com}

\author{Chloe LeGendre}
\affiliation{%
  \institution{Netflix}
  \country{USA}
}
\email{clegendre@netflix.com}

\renewcommand{\shortauthors}{Debevec and LeGendre}

\begin{abstract}
We present a technique to reduce the dynamic range of an HDRI lighting environment map in an efficient, energy-preserving manner by spreading out the light of concentrated light sources.  This allows us to display a reasonable approximation of the illumination of an HDRI map in a lighting reproduction system with limited dynamic range such as virtual production LED Stage.  The technique identifies regions of the HDRI map above a given pixel threshold, dilates these regions until the average pixel value within each is below the threshold, and finally replaces each dilated region’s pixels with the region's average pixel value.  The new HDRI map contains the same energy as the original, spreads the light as little as possible, and avoids chromatic fringing.
\end{abstract}


\begin{teaserfigure}
\vspace{-10pt}
\begin{center}
\begin{tabular}{ccc}
  \includegraphics[width=1.9in]{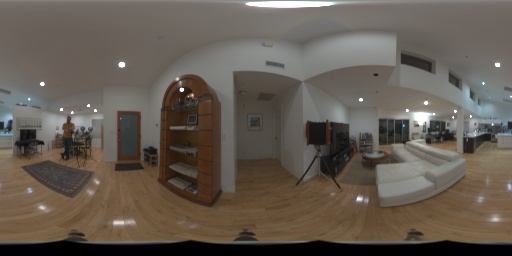} & \includegraphics[width=1.9in]{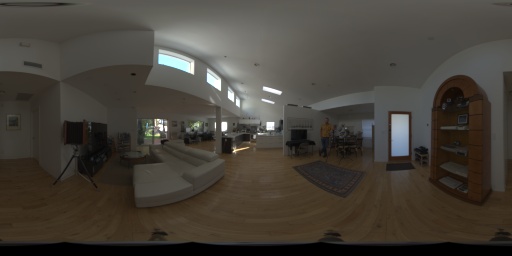} &
  \includegraphics[width=1.9in]{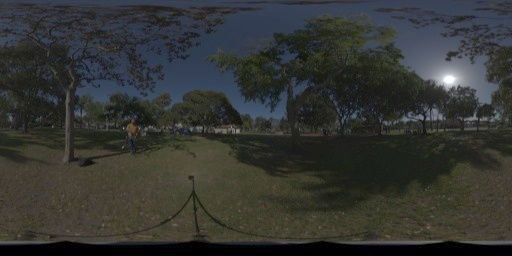} \\
  \includegraphics[width=1.9in]{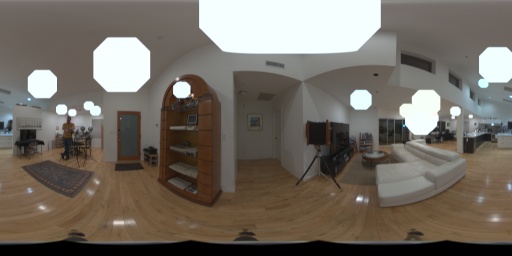} & \includegraphics[width=1.9in]{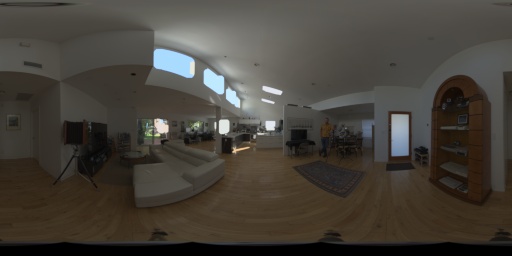} &
  \includegraphics[width=1.9in]{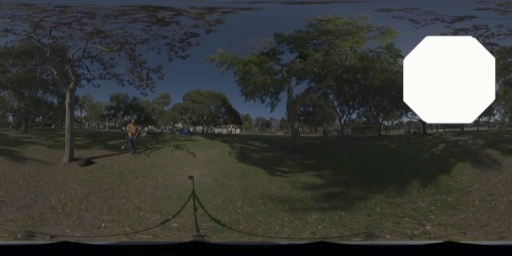} \\
\end{tabular}
  \vspace{-5pt}
  \caption{We reduce the dynamic range of HDRI maps (top) by dilating overexposed regions while preserving energy (bottom).}
  \Description{HDRI lighting environments before and after light source dilation.}
  \label{fig:teaser}
  \end{center}
\end{teaserfigure}

\maketitle
\pagestyle{plain}

\section{Background}

In a virtual production stage, an actor is surrounded by individually controllable LEDs, built of individual light sources \cite{Debevec:2002} or LED display panels \cite{Hamon:2014}. The popularity of this production technique has been in pursuit of three benefits: 1) To provide an in-camera background behind the actors which can eliminate the need for greenscreen, keying, and compositing; 2) To provide image-based lighting on the actors, illuminating them with images of the environments they are composited into; and 3) To show the actors and crew images of the scene to motivate eyelines and guide camera framing.  While current LED panels are effective for in-camera backgrounds and showing images to the case and crew, they often fall short as image-based lighting instruments due to limited dynamic range, making them unable to reproduce the brightness of concentrated light sources in a scene.  As a result, cinematographers must typically bring in additional studio lighting to compensate for this missing scene illumination, which adds expense and complexity to a shoot.

In this work, we will modify the image-based lighting environment to spread out the energy of concentrated light sources so that the scene illumination can be represented within the dynamic range of typical LED panels.

\section{Technique}

\begin{figure}[h]
\vspace{-5pt}
\begin{tabular}{cc}
\includegraphics[width=1.4in]{pano_images/ccm/HouseLightsKitchen_Panorama_8stopsup_recoveredhighlights_hdr_WHITESQEXP.jpg} & \includegraphics[width=1.4in]{pano_images/ccm/HouseLightsKitchen_Panorama_8stopsup_recoveredhighlights_hdr_WHITESQEXP_3x3_alt_plus_box.jpg} \\
\small{(a) HDRI Map} & \small{(b) LDR result} \\
\includegraphics[width=1.4in]{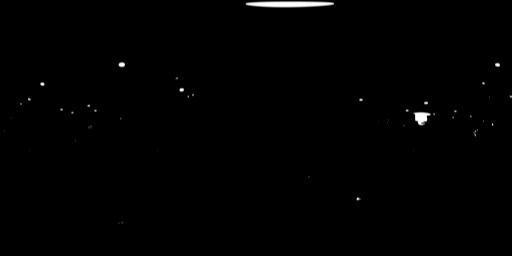} &  \includegraphics[width=1.4in]{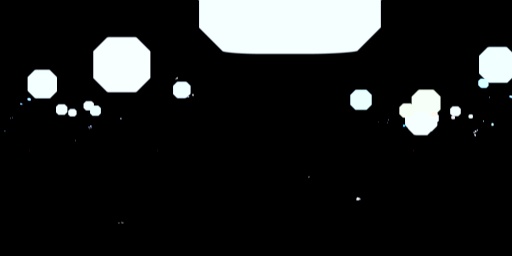} \\
\small{(c) saturated regions} & \small{(d) dilated regions} \\
\end{tabular}
\vspace{-10pt}
\caption{Saturated pixel regions are detected in the HDRI map, and dilated until the pixels no longer saturate.}
\label{fig:panos}
\end{figure}

\begin{figure*}[ht]
\vspace{-5pt}
\begin{tabular}{ccc}
\includegraphics[width=2.in]{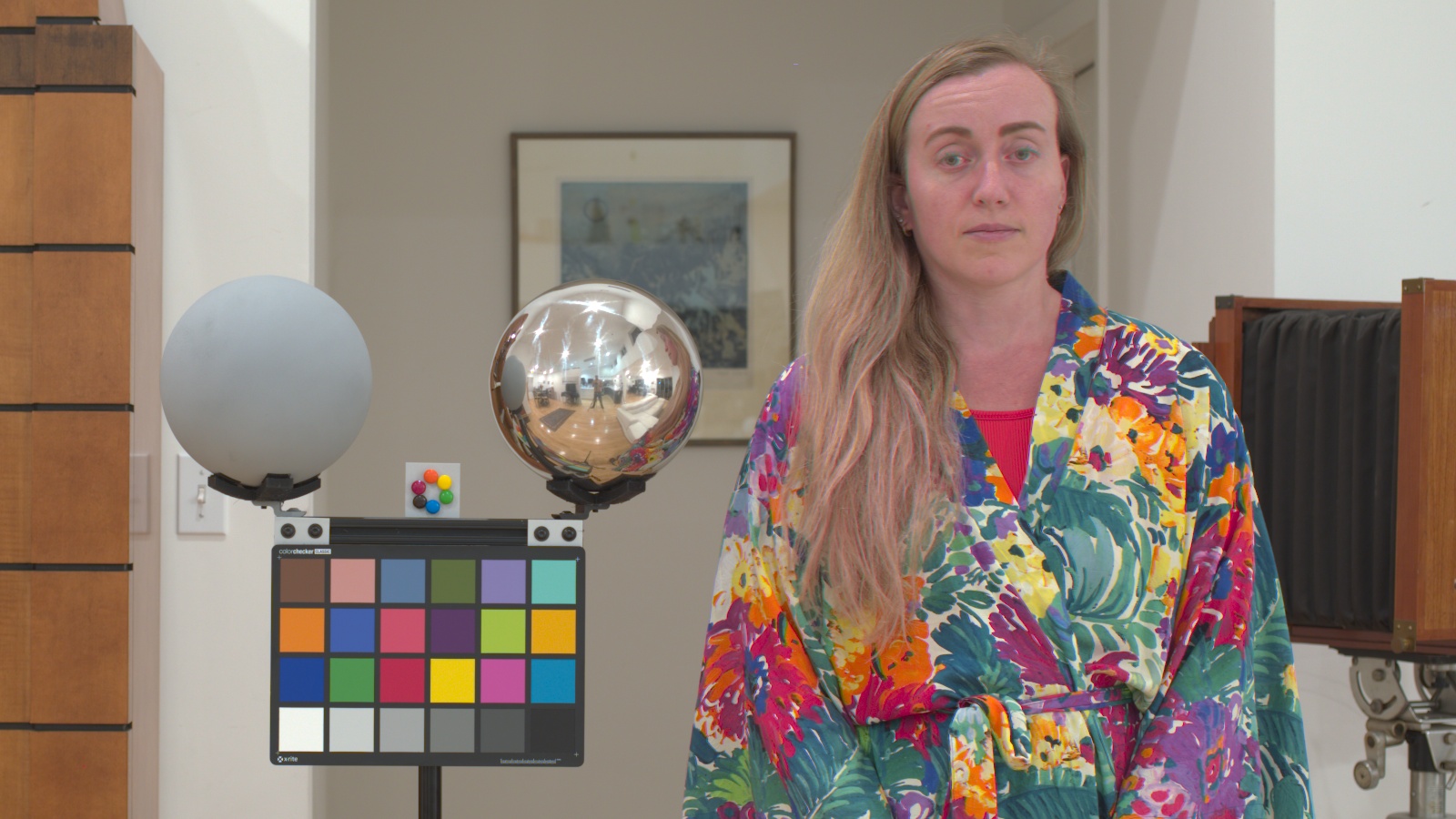} &
\includegraphics[width=2.in]{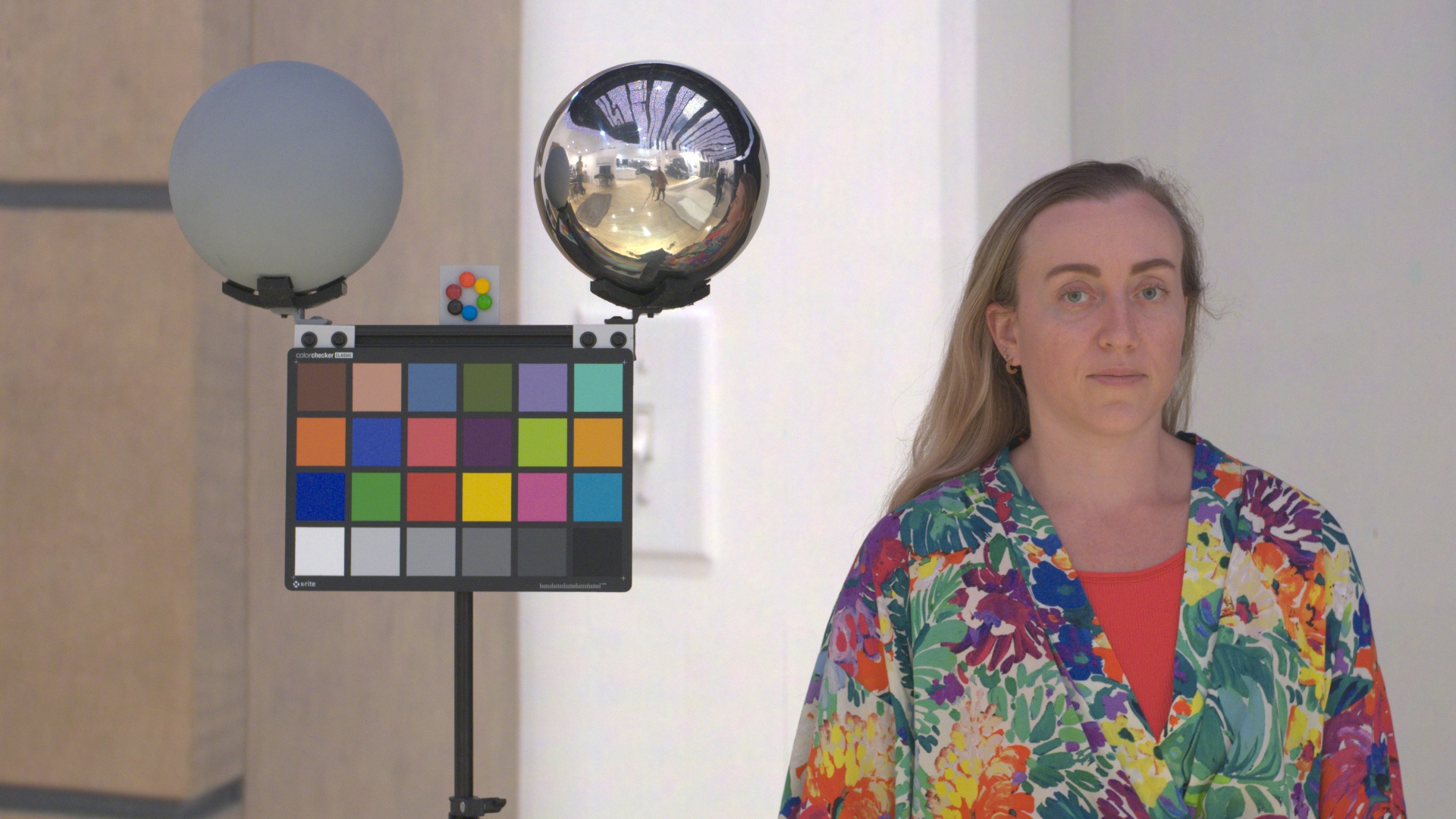} & 
\includegraphics[width=2.in]{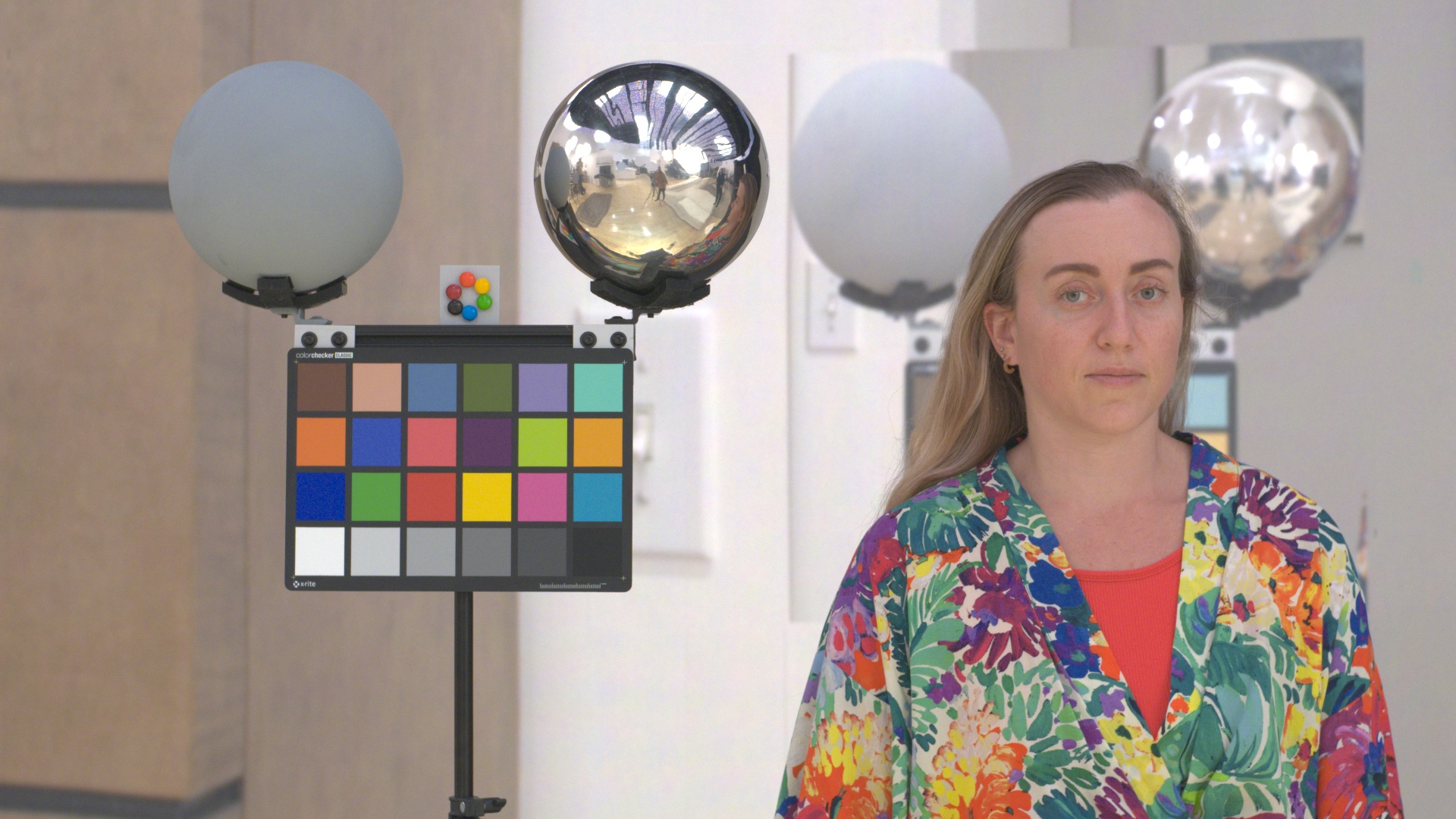} \\
\small{(a) real environment} & \small{(b) VP clipped LDR IBL} & \small{(c) VP dilated LDR IBL} \\
\includegraphics[width=2.in]{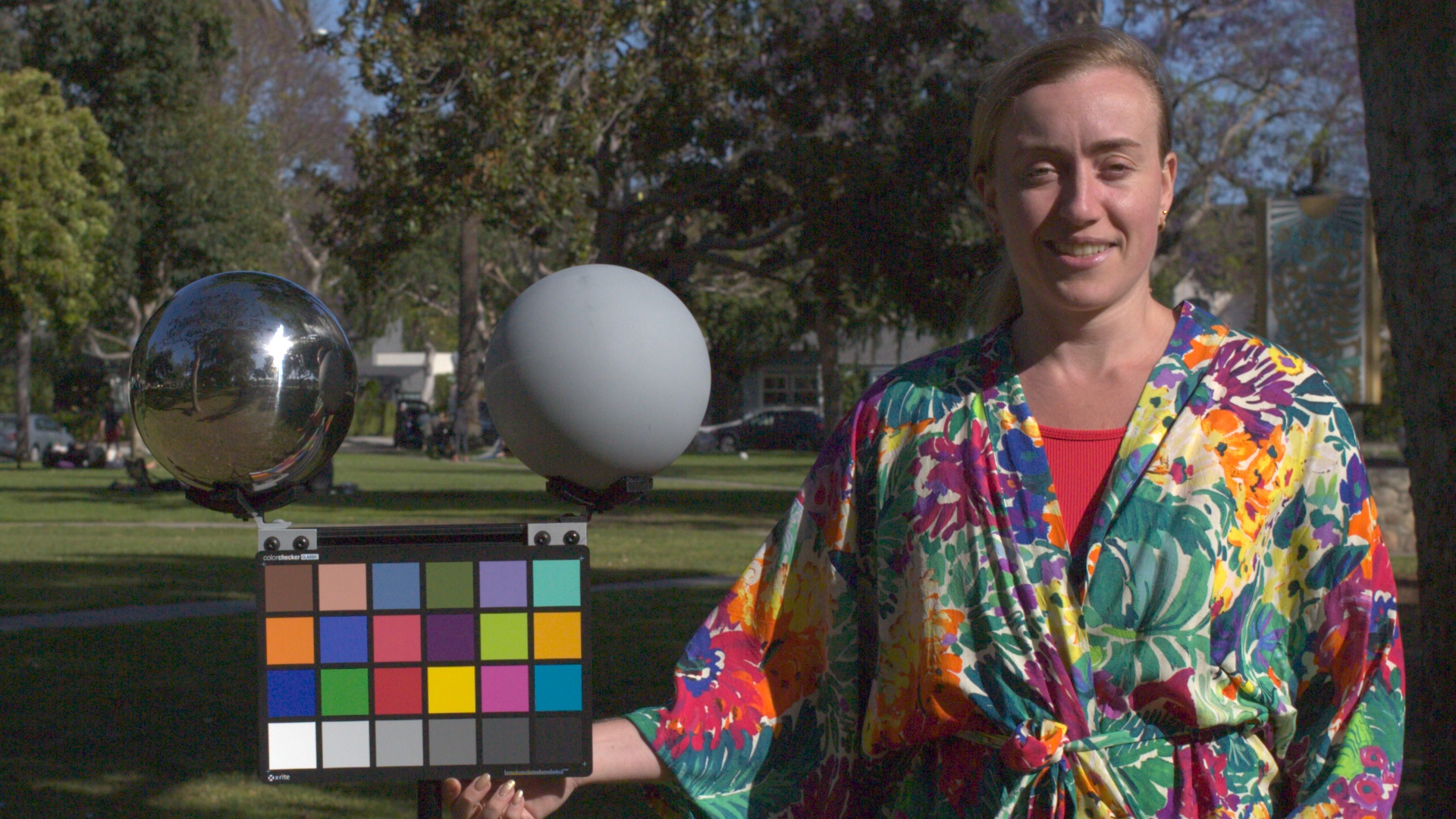} &
\includegraphics[width=2.in]{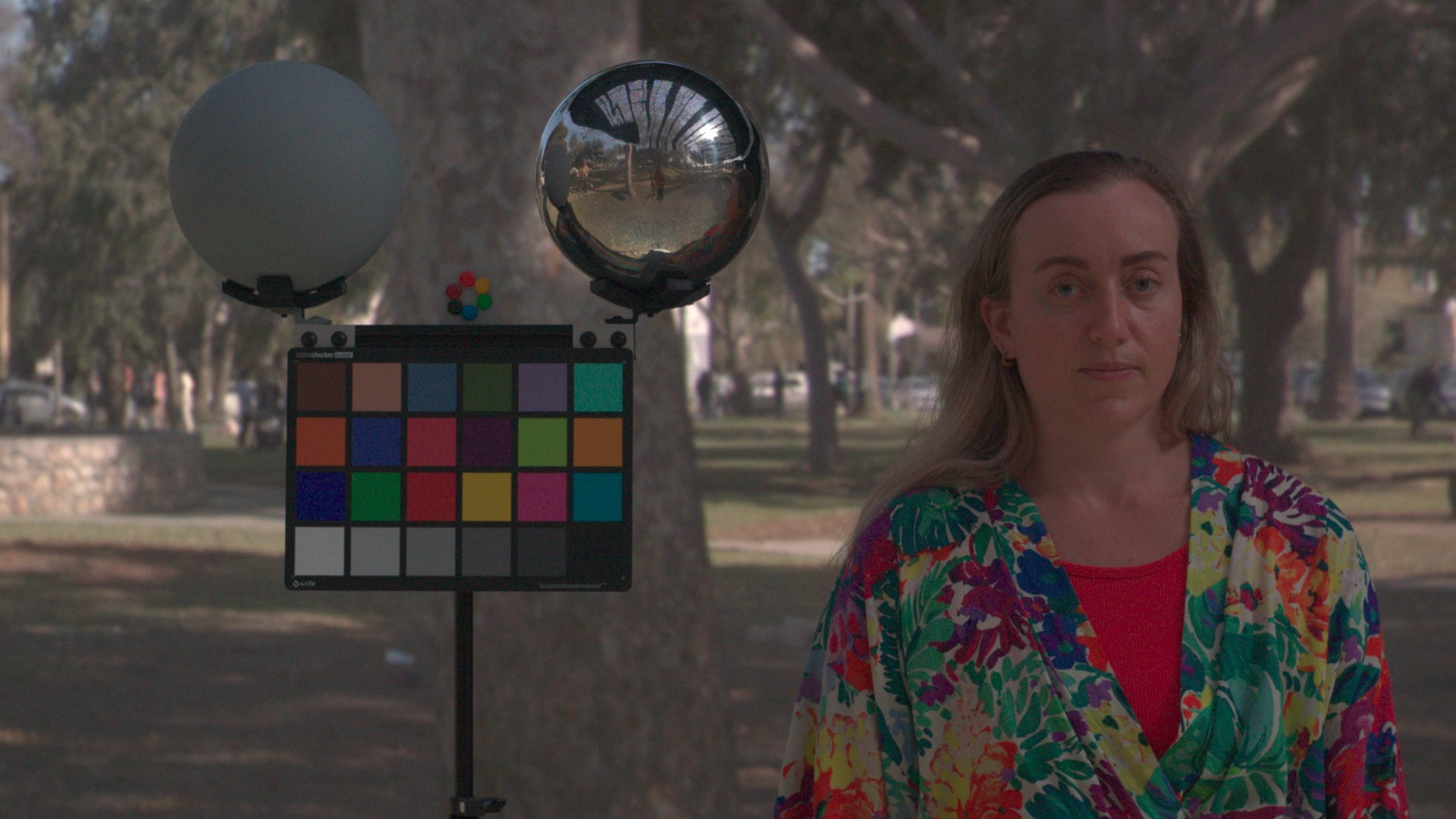} & 
\includegraphics[width=2.in]{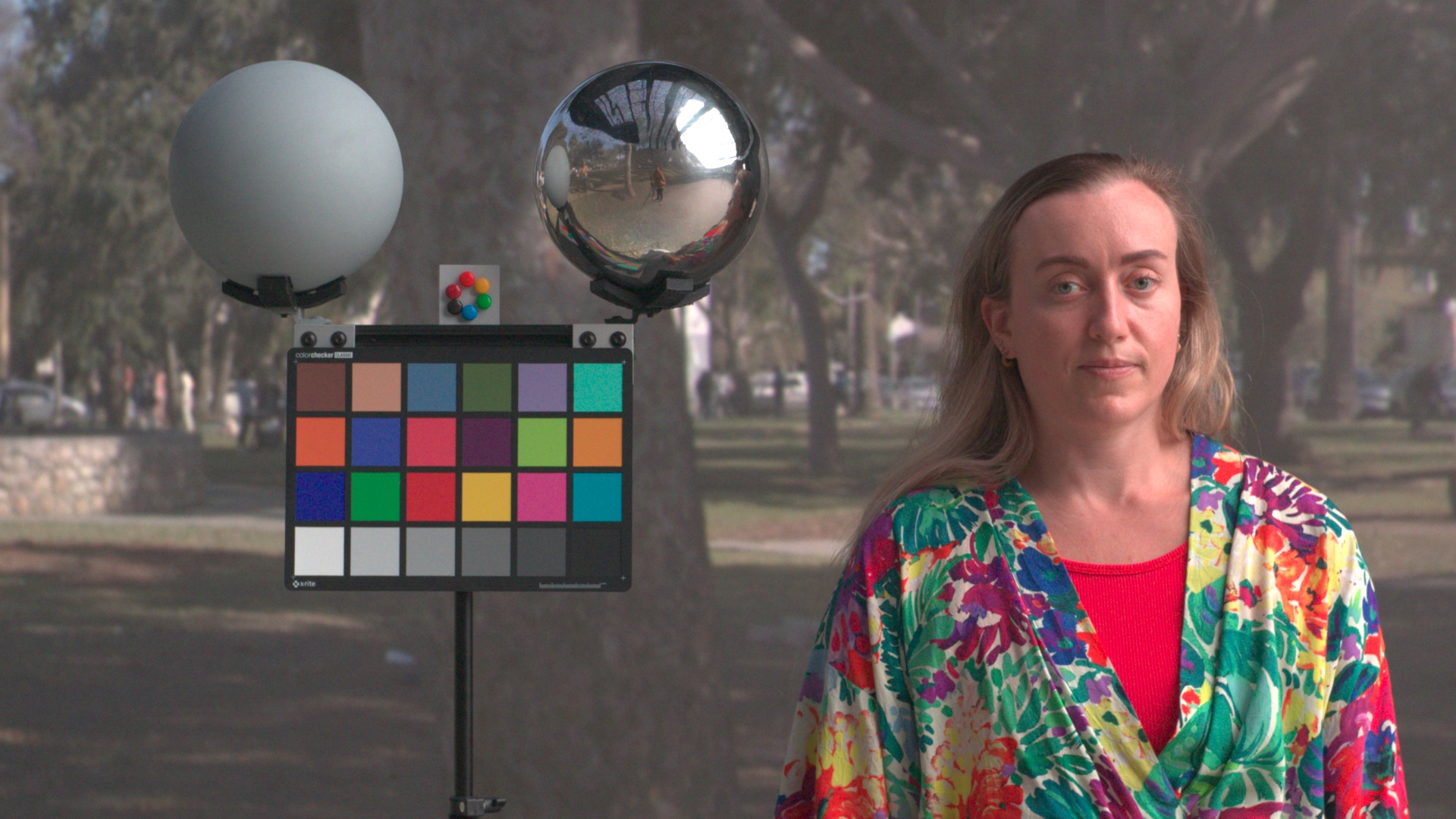} \\
\small{(d) real environment} & \small{(e) VP clipped LDR IBL} & \small{(f) VP dilated LDR IBL} \\
\end{tabular}
\vspace{-10pt}
\caption{(a, d). Subject in a real interior/outdoor environment. (b, e). In an LED stage displaying the clipped HDRI. (c, f). In the LED stage displaying the dilated environment.}
\label{fig:people_result}
\end{figure*}

\begin{figure*}[ht]
\vspace{-5pt}
\begin{tabular}{ccc}
\includegraphics[width=2.in]{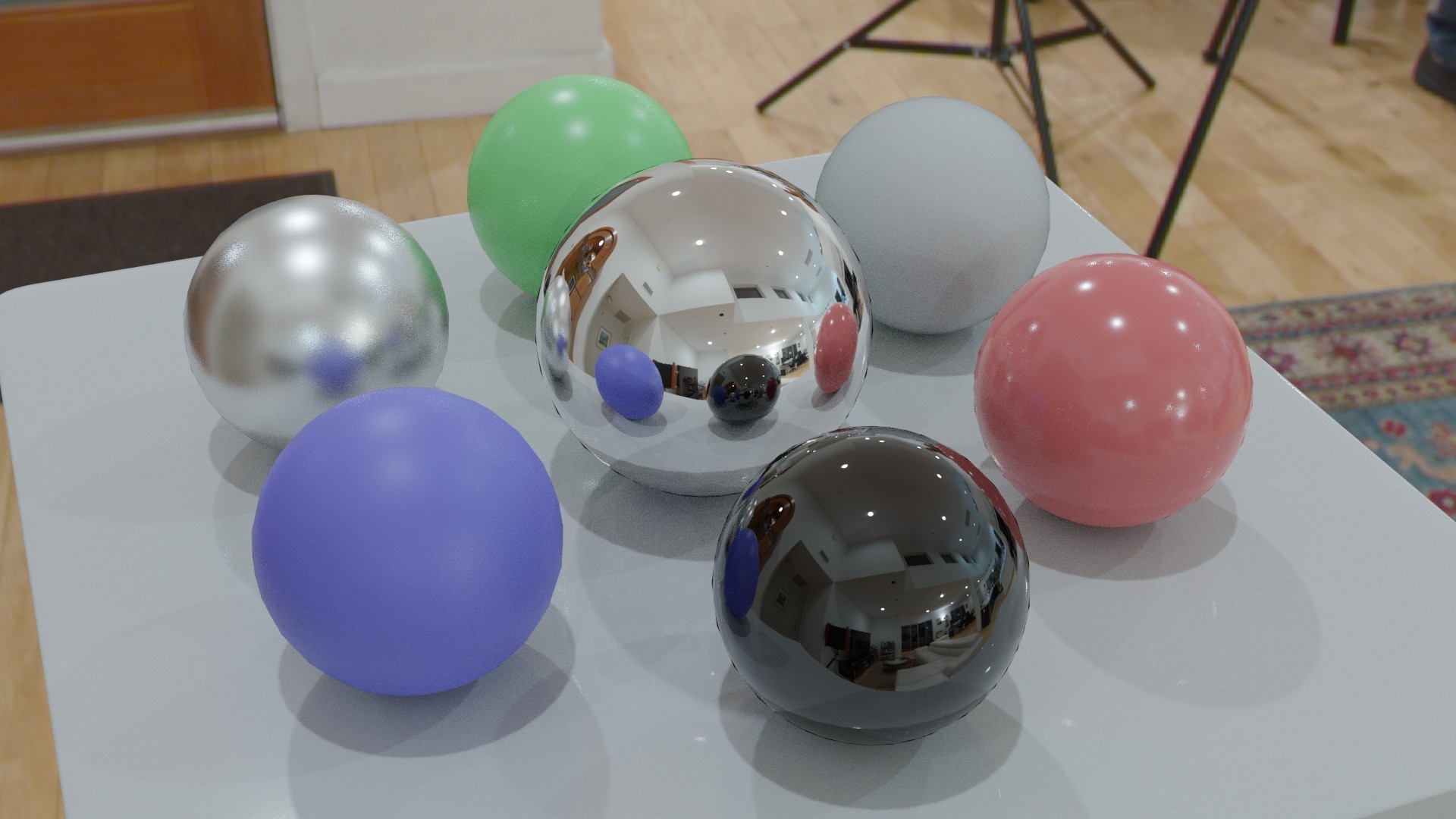} &
\includegraphics[width=2.in]{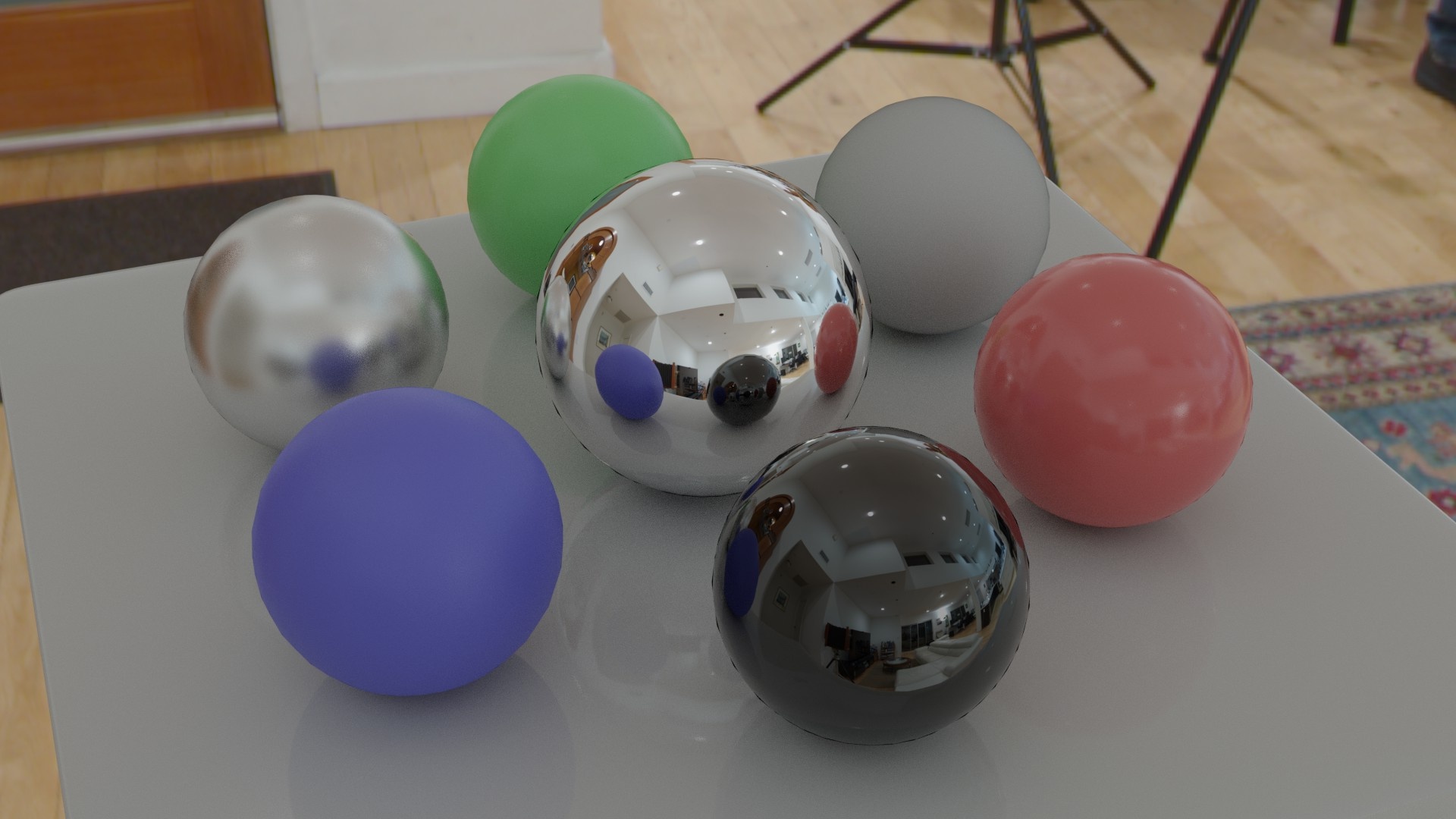} & 
\includegraphics[width=2.in]{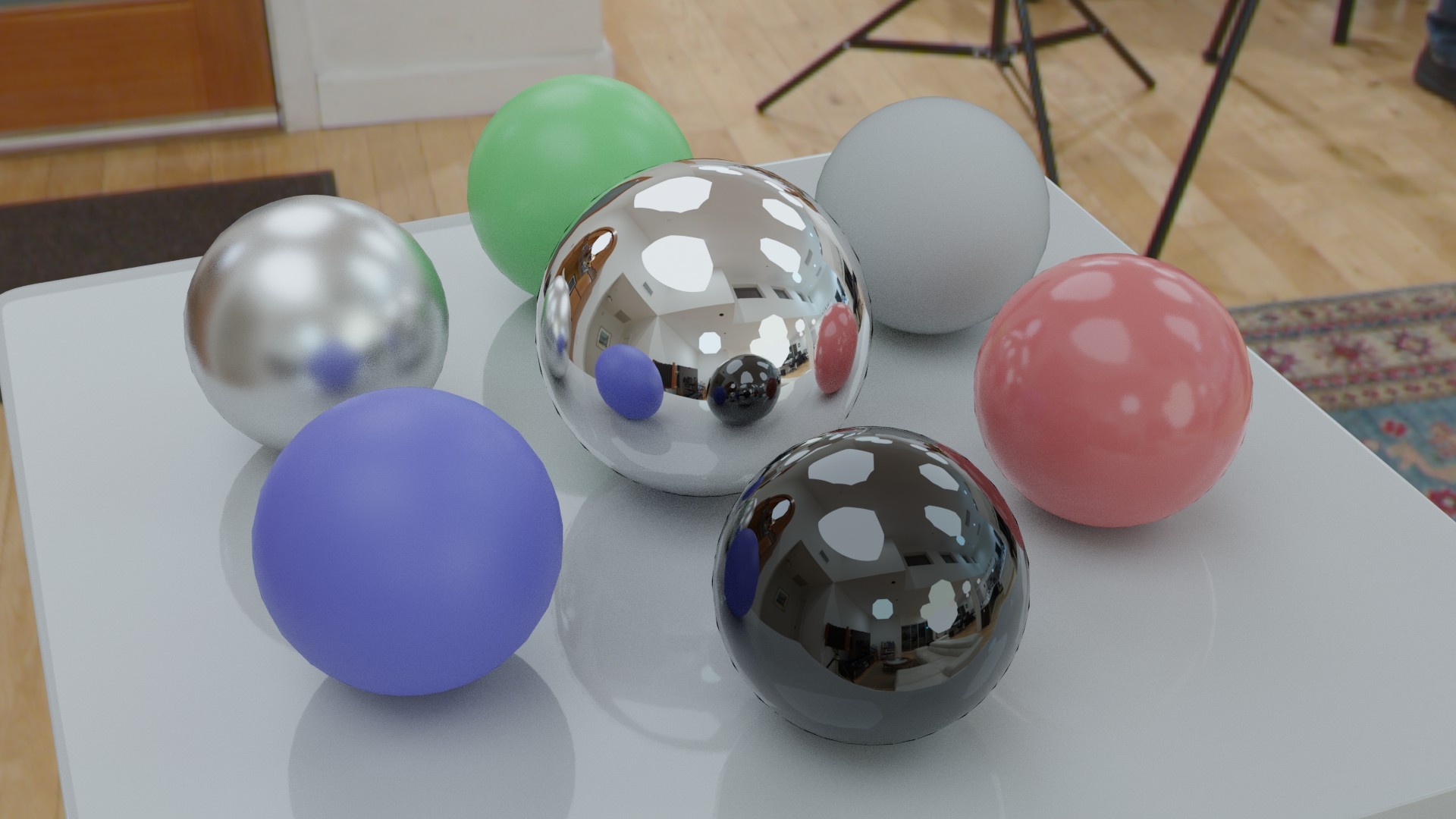} \\
\small{(a) original HDRI} & \small{(b) clipped LDR IBL} & \small{(c) dilated LDR IBL} \\
\includegraphics[width=2.in]{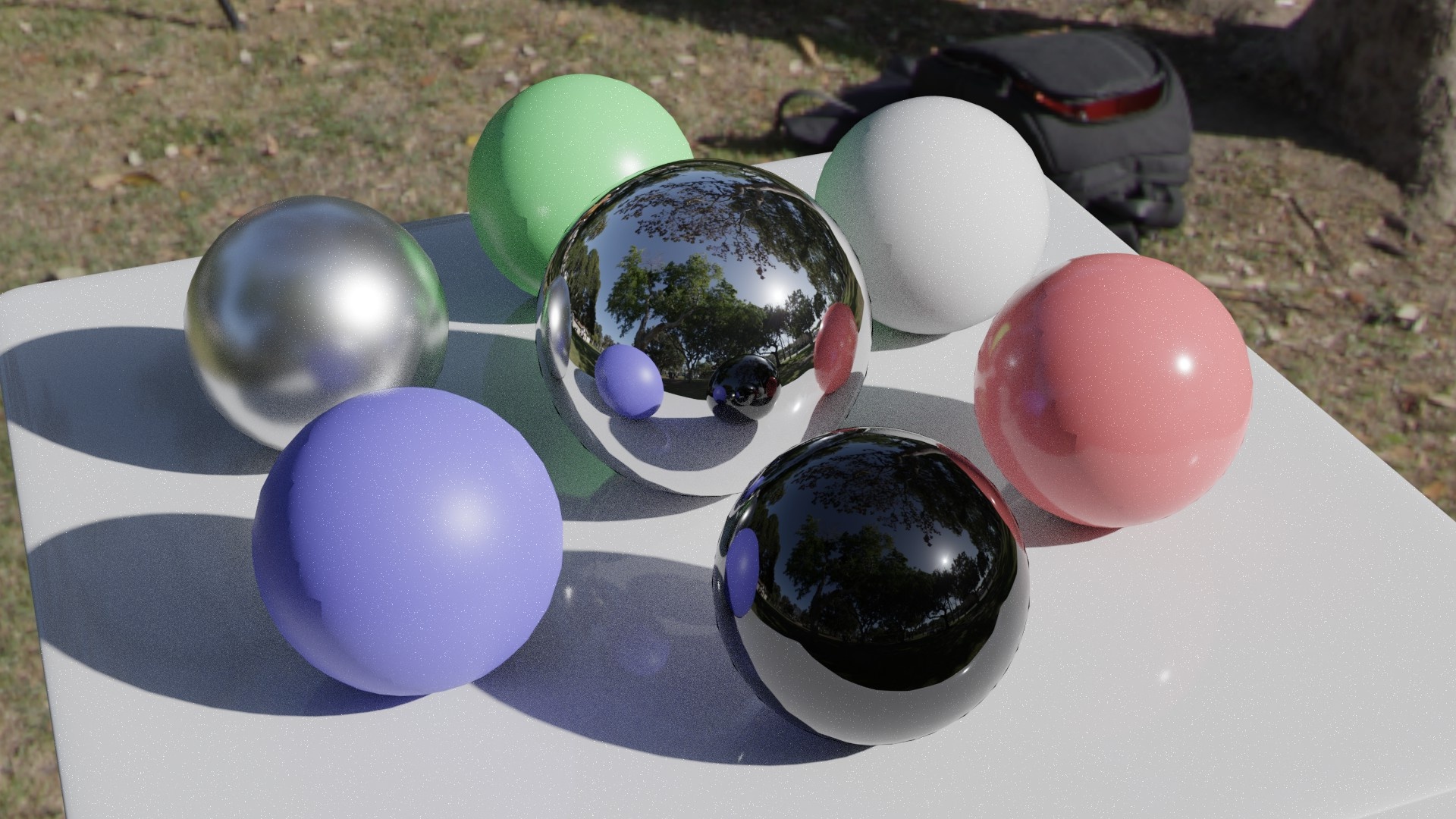} &
\includegraphics[width=2.in]{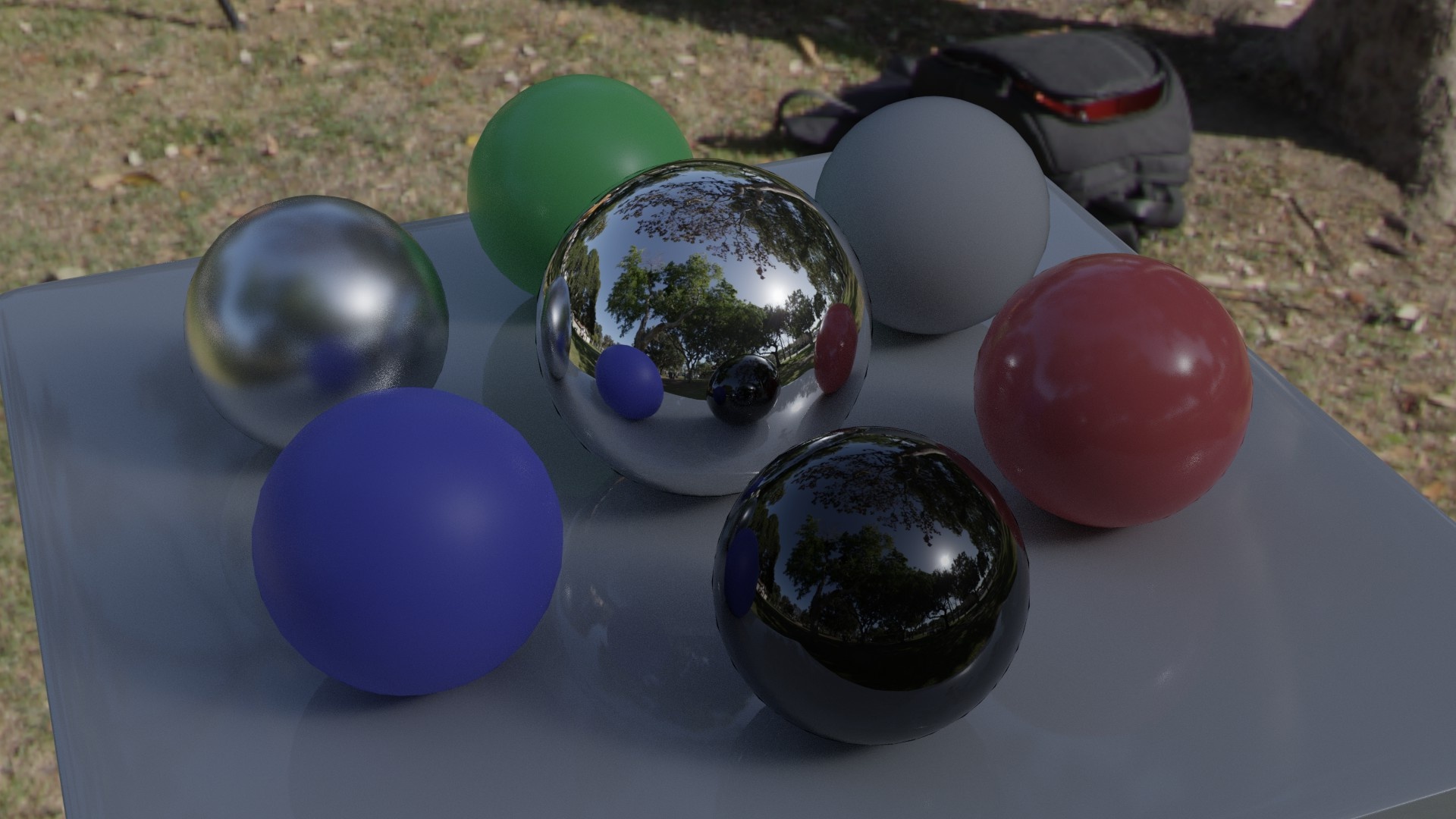} & 
\includegraphics[width=2.in]{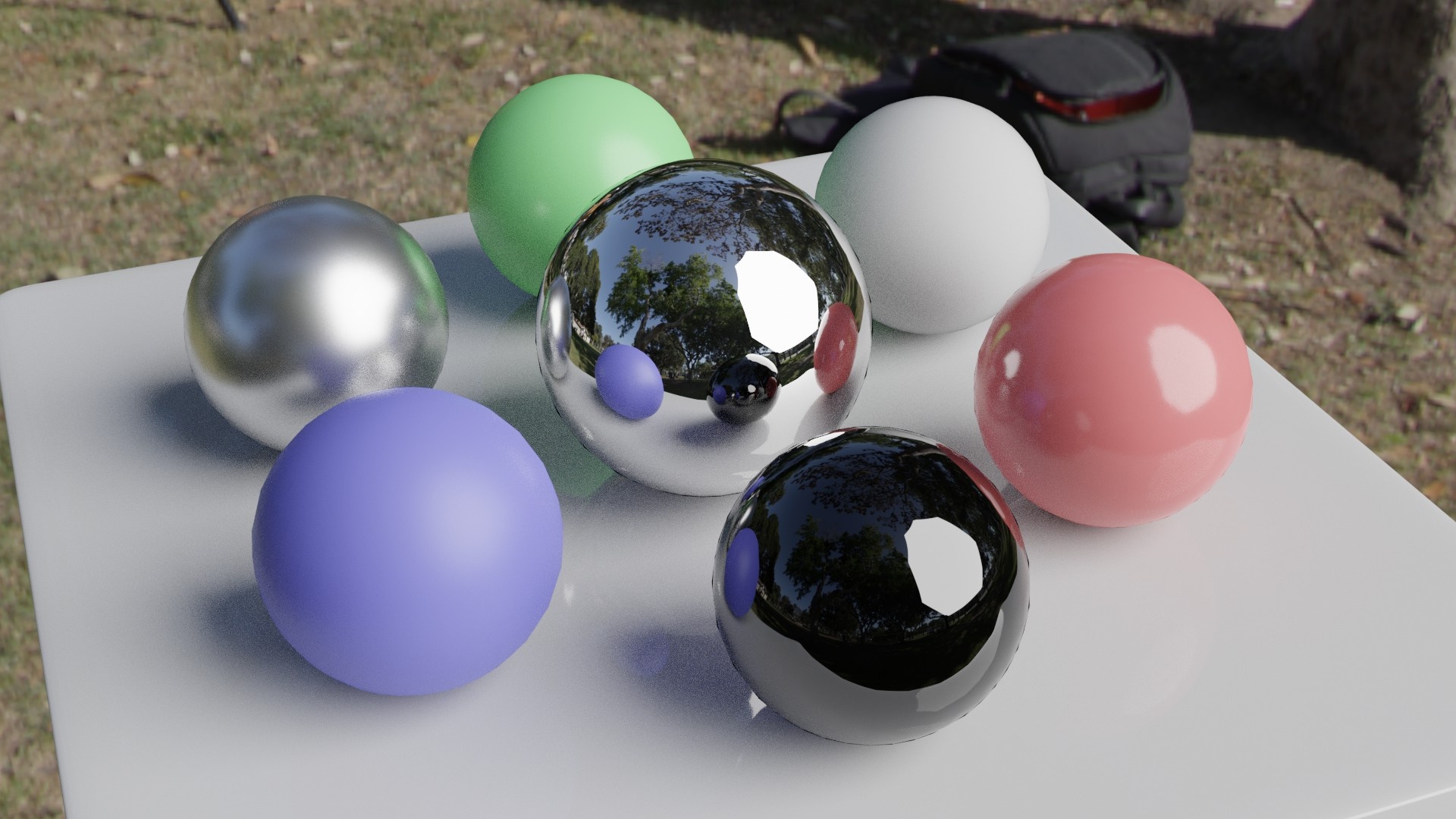} \\
\small{(d) original HDRI} & \small{(e) clipped LDR IBL} & \small{(f) dilated LDR IBL} \\
\end{tabular}
\vspace{-10pt}
\caption{Renderings using: (a, d) HDR IBL; (b, e) clipped LDR IBL; (c, f) dilated LDR IBL.}
\label{fig:ibl_result}
\end{figure*}

Our technique is simple and efficient, leveraging the fact that any area of an image-based lighting environment can be replaced with its average pixel value without changing the total energy of the environment, and changing the direction of the lighting by at most the radius of the area. Our algorithm uses the classic connected components techniques implemented in OpenCV to process the HDRI Map using this process:

\begin{enumerate}
    \item Identify pixels in the HDRI map where one channel exceeds the maximum value displayable on the LED panels, which without loss of generality we will assume to be 1.
    \item Group these saturated pixels into connected components.
    \item For each connected component (CC):
    \begin{enumerate}
     \item Compute the CC’s average pixel value $\bar{x}$.
     \item If no channel of $\bar{x}$ exceeds 1, reset the entire CC’s pixel values with $\bar{x}$.
     \item Else, dilate the perimeter of the CC by one pixel, avoiding intersections with other CC's.
     \item If any CC’s were dilated in step 3(c), repeat step 3.
    \end{enumerate}
\end{enumerate}

In practice, the new $\bar{x}$ values can be computed from the previous ones as a weighted average of the previous $\bar{x}$ and the average value of the pixels added during dilation.  Intersections are avoided by intersecting the CC’s pairwise and removing duplicated pixels from one of the regions.  If there is more energy in the HDRI map than can be spread evenly across all directions without saturating, then it will become impossible to dilate the regions without intersection, and the algorithm should terminate suggesting that the pixel values in the HDRI map should be attenuated.  The average pixel values $\bar{x}$ should be computed with respect to the solid angle subtended by each pixel, which for equirectangular maps is proportional to the cosine of the inclination, $\cos(\phi)$. Fig. \ref{fig:panos} shows this process for an interior HDRI environment.

\section{Results}

Fig. \ref{fig:panos}(a) shows an HDRI map of a living room with several ceiling lights with pixel values above 1. Fig. \ref{fig:panos}(c) shows the CC's of the clipped areas. Fig. \ref{fig:panos}(d) shows the CC's after dilation, with each colored by the average pixel value of dilated CC, now no longer saturating. Fig. \ref{fig:panos}(b) shows these dilated light source regions rendered back into the original map, forming an LDR result with the same total light energy. 

Fig. \ref{fig:people_result}(a, d) shows an actor in two real lighting environments (shown at the left and right of Fig. \ref{fig:teaser}). Fig. \ref{fig:people_result}(b, e) shows the actors on a VP stage, lit with a \textit{clipped} LDR IBL corresponding to the environment. Fig. \ref{fig:people_result}(c, f) shows the actors lit by the \textit{dilated} LDR IBL, with an reasonable match to the original lighting, somewhat diffused.

Fig. \ref{fig:ibl_result}(a, d) shows a CG scene illuminated by two of the full HDRI environments of Fig. \ref{fig:teaser}(left and right).  Fig \ref{fig:ibl_result}(b, e) shows the scene illuminated by the \textit{clipped} LDR environments, with missing illumination.  Fig \ref{fig:ibl_result}(c, f) shows the scene illuminated by the \textit{dilated} LDR environments, with softer shadows and highlights but similar appearance to the originals.

\section{Discussion and Future Work}

Our technique can transform HDRI lighting reasonably accurately into an LDR image as long as the light sources are neither too bright nor too large in the original.  While the technique softens shadows and highlights, the overall brightness, color, and directionality of the illumination is maintained.  In future work, we would like to modify the technique to address heterogeneous brightness capabilities of the wall and ceiling panels in a VP stage, and to dilate multispectral lighting environments \cite{LeGendre:2016} with more than three LED color channels.

\begin{acks}
We thank Eyeline Studios, Connie Siu, Lukas Lepicovsky, and Stephan Trojansky for their help displaying our lighting environments in a VP LED stage.
\end{acks}

\bibliographystyle{ACM-Reference-Format}
\bibliography{light_dilation}

\end{document}